\documentstyle[12pt,aasms4]{article}
\input{psfig}

\begin{document}

\title{The structure and evolution of weakly self-interacting cold dark matter halos}

\author{Andreas Burkert}
\affil{Max-Planck-Institut f\"ur Astronomie, K\"onigstuhl 17,\\
       D-69117 Heidelberg, \\Germany}
\authoremail{burkert@mpia-hd.mpg.de}

\begin{abstract}
The evolution of halos consisting of weakly self-interacting
dark matter particles is investigated using a new numerical Monte-Carlo
N-body method. The halos initially
contain kinematically cold, dense $r^{-1}$-power-law cores.
For interaction cross sections 
$\sigma^* = \sigma/m_p \geq$ 10 - 100 cm$^2$ g$^{-1}$
weak self-interaction leads to the formation of
isothermal, constant density cores within a Hubble time as a result
of heat transfer into the cold inner regions.
This core structure is in good agreement with the 
observations of dark matter rotation curves
in dwarf galaxies. The isothermal core radii and core densities
are a function of the halo scale radii and scale masses 
which depend on the cosmological model. 
Adopting the currently popular  $\Lambda$CDM model, the predicted
core radii and core densities are in good agreement with the observations.
For large interaction cross sections, massive dark halos
with scale radii r$_s \geq 1.4 \times 10^4$ (cm$^2$ g$^{-1}$ /$\sigma^*$) kpc could
experience core collapse during their lifetime,
leading to cores with singular isothermal density profiles.
\end{abstract}

\keywords{dark matter -- galaxies: halos -- galaxies: formation
-- galaxies: kinematics and dynamics -- methods: numerical}

\section{Introduction}
Cosmological models with a dominant cold dark matter component predict
dark matter halos with strongly bound, kinematically 
cold cores (Dubinski \& Carlberg 1991, Warren {\it et al.} 1992,  
Navarro {\it et al.} 1997). Within the core region, the dark matter
density increases as a power-law $\rho \sim r^{-\gamma}$ with $\gamma$ 
in the range of 1 to 2 and the velocity dispersion $\sigma$ decreases 
towards the center (Carlberg 1994, Fukushige \& Makino 1997). Numerous numerical 
simulations (e.g. Moore {\it et al.} 1998, Huss {\it et al.} 1999, Jing \& Suto 2000),
as well as analytical theory (Syer \& White 1998, Kull 1999), have shown that such
a core structure follows naturally from
collisionless hierarchical merging of cold dark matter halos, 
independent of the adopted cosmological parameters. 

It has recently become clear that on galactic scales the predictions of cold 
dark matter models are not in agreement with several observations. 
High-resolution calculations by 
Klypin {\it et al.} (1999) and Moore {\it et al.} (1999) have shown that the
predicted number and mass distribution of galaxies in galactic clusters is consistent 
with the observations. However, on scales of the Local Group, 
roughly one thousand dark matter halos should exist as separate, self-gravitating objects,
whereas less than one hundred galaxies are observed. This disagreement can be
attributed to the high core densities of satellite dark halos in cosmological models
which stabilize them against tidal disruption on galactic scales.
Mo {\it et al.} (1998) and lateron Navarro and Steinmetz (2000) found
that cold-dark matter models reproduce well the I-band Tully-Fisher slope and 
scatter. They however fail to match the zero-point of the Tully-Fisher relation as well as
the relation between disk rotation speed and angular momentum. 
Again, this problem can be traced to the excessive central concentrations
of cold dark halos. Finally, recent observations of dark matter dominated rotation 
curves of dwarf galaxies have indicated shallow dark matter cores which can
be described by isothermal spheres with finite central densities 
(Moore 1994, Flores \& Primack 1994, Burkert 1995, 
de Blok \& Mc Gaugh 1997, Burkert \& Silk 1999, Dalcanton \& Bernstein 2000,
see however van den Bosch {\it et al.} 1999), in contrast to the power-law cusps,
expected from cosmological models.  The disagreement between observations
and theory indicates that a substantial revision to the cold dark matter scenario 
might be required which could provide valuable insight into the origin 
and nature of dark matter.

Motivated by these problems, Spergel \& Steinhardt (1999) proposed a model where
dark matter particles experience weak self-interaction on scales of kpc to Mpc for 
typical galactic densities. They noted that self-interaction could lead to satellite 
evaporation due to the dark particles within the satellites being kicked out by 
high-velocity encounters with dark particles from the surrounding dark halo of the parent 
galaxy. In order for weak interaction to be important on galactic scales,
they estimate that the ratio of the collision cross section 
and the particle mass should be of order $\sigma_{wsi}/m_p \approx$ 1 cm$^2$ g$^{-1}$.

The Spergel and Steinhardt model has already motivated several follow-up studies.
For example, Ostriker (1999) demonstrated that weak self-interaction would
have the interesting side product of naturally growing black holes with masses
in the range $10^6 - 10^9$ M$_{\odot}$ in galactic centers. 
Hogan \& Dalcanton (2000) investigated
analytically the effect of particle self-interactions on the structure and
stability of galaxy halos. 
Moore et al. (2000), adopting a gas-dynamical approach, showed that in
the limit of infinitely large interaction cross sections dark halos would develop
singular isothermal density profiles which are not in agreement with observations.
Mo \& Mao (2000) and Firmani et al. (2000) investigated the affect of self-interaction
on rotation curves. In addition, models of repulsive dark matter (Goodman 2000),
fluid dark matter (Peebles 2000) and self-interacting warm dark 
matter (Hannestad \& Scherrer 2000) have recently been discussed.

In this paper we will investigate the effect of weak self-interaction
on the internal structure of cold dark matter halos. If the interaction cross section
is not exceptionally large, the dark matter system cannot be treated
as a collision dominated, hydrodynamical fluid. Section 2 therefore 
introduces a new numerical Monte-Carlo-N-body (MCN) method for
weakly interacting particle systems. Initial conditions are discussed in section 3.
Using the MCN-method, the evolution of weakly self-interacting dark matter
halos is investigated in section 4.  Conclusions follow in section 5.

\section{The Monte-Carlo N-body method}

Within the framework of weak self-interacting, the mean free path $\lambda$
of a dark matter particle is determined by 
$\lambda = (\rho \sigma^*)^{-1}$,
where $\rho$ is the local dark matter mass density
and $\sigma^* = \sigma_{wsi}/m_p$ is the ratio between the self-interaction collision 
cross section $\sigma_{wsi}$ and the particle mass $m_p$. 
If, within a timestep $\Delta t$,
the path length $l = v \Delta t$ of a particle with velocity v is short compared to 
$\lambda$,  the probability $P$ for it to interact with another particle
can be approximated by
 
\begin{equation}
P = l/\lambda = \sigma^* \rho v \Delta t.
\end{equation}

We use a Monte-Carlo approach in order to include weak self-interaction
in a collisionless N-body code that
utilizes the special purpose hardware GRAPE (GRAvity PipE; Sugimoto et al. 1990)
in order to determine the gravitational forces between the dark matter 
particles by direct summation. For each particle, a list of its 50 
nearest neighbors is returned by the boards which allows the determination
of the local dark matter mass density $\rho$.
The particle experiences an interaction with its nearest neighbor with a probability 
given by equation (1). Each weak interaction changes the
velocities of the two interacting particles.
Here, due to the lack of a more sophisticated theory, we assume that the 
interaction cross section is isotropic and that the interaction is 
completely elastic. In this case, the directions of the velocity vectors 
after the interaction are randomly chosen and their absolute values are 
completely determined by the requirement of energy and momentum conservation.

The computational timestep $\Delta t$ must be chosen 
small enough in order to guarantee that the evolution is independent of the 
numerical parameters, that is the adopted timestep and the number of particles. 
Otherwise, particles with large velocities could penetrate too deeply into 
a dense region like the core of a dark matter halo, violating the 
requirement $l << \lambda$. Test calculations 
have shown that $\Delta t \leq \eta (\sigma^* \rho v)^{-1}$ with
$\eta \approx 0.1$  leads to reliable results 
that are independent of the numerical parameters. 

\section{Initial conditions}

Cold dark matter halos form on dynamical timescales. If 
$\sigma^*$ is small enough, the halos will achieve an equilibrium
state within  a few dynamical timescales that is determined by collisionless 
dynamics alone, before self-interaction becomes
important. The structure of the halos subsequently changes due to self-interaction
on longer timescales. This secular evolution is similar to the
long-term evolution of globular clusters
which experience core collapse due to gravitational 2-body encounters
after virialization.
 
We start with an equilibrium model of a virialized 
dark matter halo and study its secular dynamical evolution due to weak self-interaction
using the MCN-method.
As initial condition, a Hernquist halo model (Hernquist 1990) is adopted.
Its density distribution is $\rho(r) = \rho_{s}/\left(r/r_s (1+r/r_s)^3 \right)$
where $\rho_{s}$ and $r_s$ are the scale density  and scale radius, respectively. 
The mass profile is $M(r)=M r^2/(r_s+r)^2$ with $M$ the finite total halo mass.
Assuming hydrostatic equilibrium and an isotropic velocity distribution, the velocity 
dispersion is zero at the center and increases outwards, reaching a maximum
at the inversion radius $r_i = 0.33 r_s$ outside of which it decreases again. A similar structure
is seen in cosmological simulations (Carlberg 1994, Fukushige \& Makino 1997).
In general, within the interesting region $r \leq r_s$ the Hernquist model 
provides an excellent fit to
the structure of cold dark matter halos that result from high-resolution cosmological models.
Only in the outermost regions do the dark matter halo profiles deviate significantly from the
Hernquist model, predicting a density distribution that decreases as $r^{-3}$ and a
dark halo mass that diverges logarithmically (Navarro {\it et al.} 1997).
Note, that our model neglects any clumpy substructure that might exist within 
dark matter halos (Moore {\it et al.} 1999). 
This should be a reasonable approximation for the 
inner regions where satellites are efficiently disrupted by tidal forces.
The evolution of weakly self-interacting, clumpy dark halos 
will be presented in a subsequent paper (see also Moore et al. 2000).

In the following, we will adopt dimensionless units: G=1, $r_s=1$ and $M=1$. 
The total mass and the mean mass density within the inversion radius
$r_i$ is $M_i=0.06$
and  $\rho_i = 0.4$, respectively, leading to a dynamical timescale within $r_i$ of 
$\tau_{dyn} = 0.8$. Most numerical calculations have been performed adopting 80000
particles and a gravitational softening length of $\epsilon$ = 0.002$\times$r$_s$. 
Test calculations
with 120000 particles did not change the results.
N-body calculations without weak interaction
have shown that the dark halo is stable and its density distribution does not
change outside of r $\geq$ 0.006$\times$r$_s$ within 20 dynamical timescales.

\section{The evolution of weakly self-interacting dark halos}

Figure 1 shows the evolution of the dark matter density distribution and the velocity
dispersion profile inside the core region, adopting a collision cross section 
$\sigma^*$=10$\times$r$_s^2$/M$_s$. The density distribution initially has
the characteristic power-law cusp and the velocity dispersion decreases towards
the center for $r < r_i$.  Within this region, the kinetic temperature inversion
leads to heat conduction inwards.  The central velocity dispersion 
increases with time and the core expands, resulting in a shallower density distribution.
After 3 dynamical timescales, an isothermal, constant density core has formed
with a radius that is of order the initial inversion radius $r_{i}$.
Subsequently, weak interactions between the kinematically hotter core and the cooler
envelope lead to a flow of kinetic energy outwards which causes the isothermal core
to contract and heat up further due to its negative specific heat, 
starting a core collapse phase. The calculations are stopped after 16 dynamical timescales
when the central density and the central velocity dispersion has increased
further by a factor
of 4 and 1.4, respectively. Note that during the core collapse phase, the system maintains an
isothermal, constant density core with the core radius decreasing with time.
Overall, the evolution of the dark halo is very similar to the secular evolution of
particle systems with Hernquist profiles that are affected by
gravitational 2-body interactions (Heggie {\it et al.} 1994, Quinlan 1999).

Several calculations with different interaction cross sections $\sigma^*$ have been
performed. In all cases, the evolution is similar to that shown in Fig. 1, independent of
the adopted collision cross section. The timescale $\tau_{iso}$ 
for the formation of the isothermal constant density core does however
depend on $\sigma^*$ with 

\begin{equation}
\tau_{iso} \approx \frac{30 \tau_{dyn}}{\sigma^*} \frac{r_s^2}{M_s} .
\end{equation}

\noindent In agreement with the
calculations of Quinlan (1999) the core collapse timescales are
roughly an order of magnitude larger than $\tau_{iso}$.

Observations of dark matter dominated dwarf galaxies show a characteristic
dark matter core structure that can be fitted well by the empirical density
distribution (Burkert 1995)
$\rho = \rho_0*(r+r_0)^{-1}(r^2+r_0^2)^{-1}$ where $\rho_0$ and $r_0$
are the isothermal core density and radius, respectively. Figure 2 compares this 
profile (solid line) with the core structure of weakly interacting dark halos at
$t=0$ (dashed line) and after core expansion at $t = \tau_{iso}$ (points with
error bars). It is well known that power-law cores do not provide a good
fit to the observations. An excellent agreement can however be achieved after 
core expansion if one adopts the following core parameters

\begin{eqnarray}
r_0 \approx 0.6 r_i \\ \nonumber
\rho_0 \approx 1.54 M_i r_i^{-3} 
\end{eqnarray}

\noindent where $M_i$ is the initial dark matter mass inside the inversion radius r$_i$.

\section{Conclusions}
The previous MCN calculations have shown that
isothermal cores with shallow density profiles form naturally in weakly
interacting dark halos. The cores have
density distributions that are in excellent agreement with the observations
of dark matter rotation curves in dwarf galaxies. 
The core size is determined by the radius r$_i$ inside which heat is conducted inwards,
that is where the initial velocity dispersion decreases towards the center. Note,
that this conclusion should be valid, independent of whether the density diverges
as $\rho \sim r^{-1}$ or even steeper 
(Moore {\it et al.} 1998, Jing \& Suto 2000) for  r $\ll r_s$.

A quantitatively comparison with the observations requires the
determination of the typical scale parameters r$_s$ and M$_s$ for dark matter halos. 
Recent cosmological $\Lambda$CDM models (Navarro \& Steinmetz 2000) 
predict that cold dark matter halos with total masses
M$_{200} \approx 10^{10} - 10^{12}$ M$_{\odot}$ should have concentrations
c = r$_{200}$/r$_{s} \approx 20$, where
M$_{200}$ is the total dark matter mass within the virial
radius r$_{200}$ which denotes the radius inside which the averaged
overdensity of dark matter is 200 times the critical density of the universe.
Adopting a Hubble constant h=0.7 leads to 
r$_{200}$ = 0.02 (M$_{200}$/M$_{\odot})^{1/3}$ kpc $\approx$ 40 -- 200 kpc
and with c=20 to scale radii r$_s \approx$ 2 -- 10 kpc.
For a NFW-profile (Navarro {\it et al.} 1997) the dark matter mass inside r$_s$ 
is M$_s \approx$ 0.1 M$_{200}$ and the core density is 
M$_s$/r$_s^3 \approx$ 0.01 M$_{\odot}$ pc$^{-3}$. 
In contrast to the Hernquist model with $r_i = 0.33 r_s$, the inversion radius
of the NFW-profiles coincides with the scale radius $r_i = r_s$ due to the
shallower outer density distribution. According to equation 3, weak interaction in NFW-halos
should therefore lead to isothermal cores with 
radii r$_0 \approx$ 0.6 r$_s \approx$ 1.2 -- 6 kpc and densities 
$\rho_0 \approx$ 1.55 M$_s$ r$_s^{-3} \approx 1.5 \times 10^{-2}$ M$_{\odot}$ pc$^{-3}$. 
The observations indicate core radii r$_0 \approx$ 2 -- 10 kpc with core densities 
$\rho_0 \approx$ 0.01 M$_{\odot}$ pc$^{-3}$
(Burkert 1995), in very good agreement with the theoretical predictions.  

In order for dark matter cores to be affected by weak self-interaction, the
core expansion timescale must be smaller than the age $\tau$ of the halo:
$\tau_{iso} \leq \tau \approx 100 \tau_{dyn}$. With equation (2) and adopting 
M$_s$/r$_s^3 \approx 0.01 $M$_{\odot}$ pc$^{-3}$,
this requirement leads to a minimum value of the collision cross section
for weak self-interaction to be important

\begin{equation}
\sigma^* \geq 100 \left( \frac{kpc}{r_s}  \right) \left(\frac{cm^2}{g} \right)
\end{equation}

\noindent Note, that this lower limit would be a factor of 25 larger
if cosmological models underestimate the scale radii of dark halos by a factor of 5.

Dark matter halos with $\tau_{iso} \approx \tau_{dyn}$ are likely to have gone
through core collapse if their ages are $\tau >> \tau_{dyn}$. This condition
requires the halo core radii to be larger than
r$_s > 1.4 \times 10^4$ (cm$^2$ g$^{-1}$)/$\sigma^*$ kpc, indicating that more massive
halos could have experienced core collapse while lower mass halos could
still be in the process of core expansion.

\acknowledgments

I would like to thank Matthew Bate for providing a subroutine to find nearest
neighbors using GRAPE and  Paul Steinhardt, Ben Moore
and Jerry Ostriker  for interesting discussions and the referee for important comments. 
Special thanks to David Spergel for
pointing out the different dependence of the inversion radius on the scale radius
for NFW- and Hernquist profiles.

\newpage

\begin{figure}[p]
\centerline{\psfig{figure=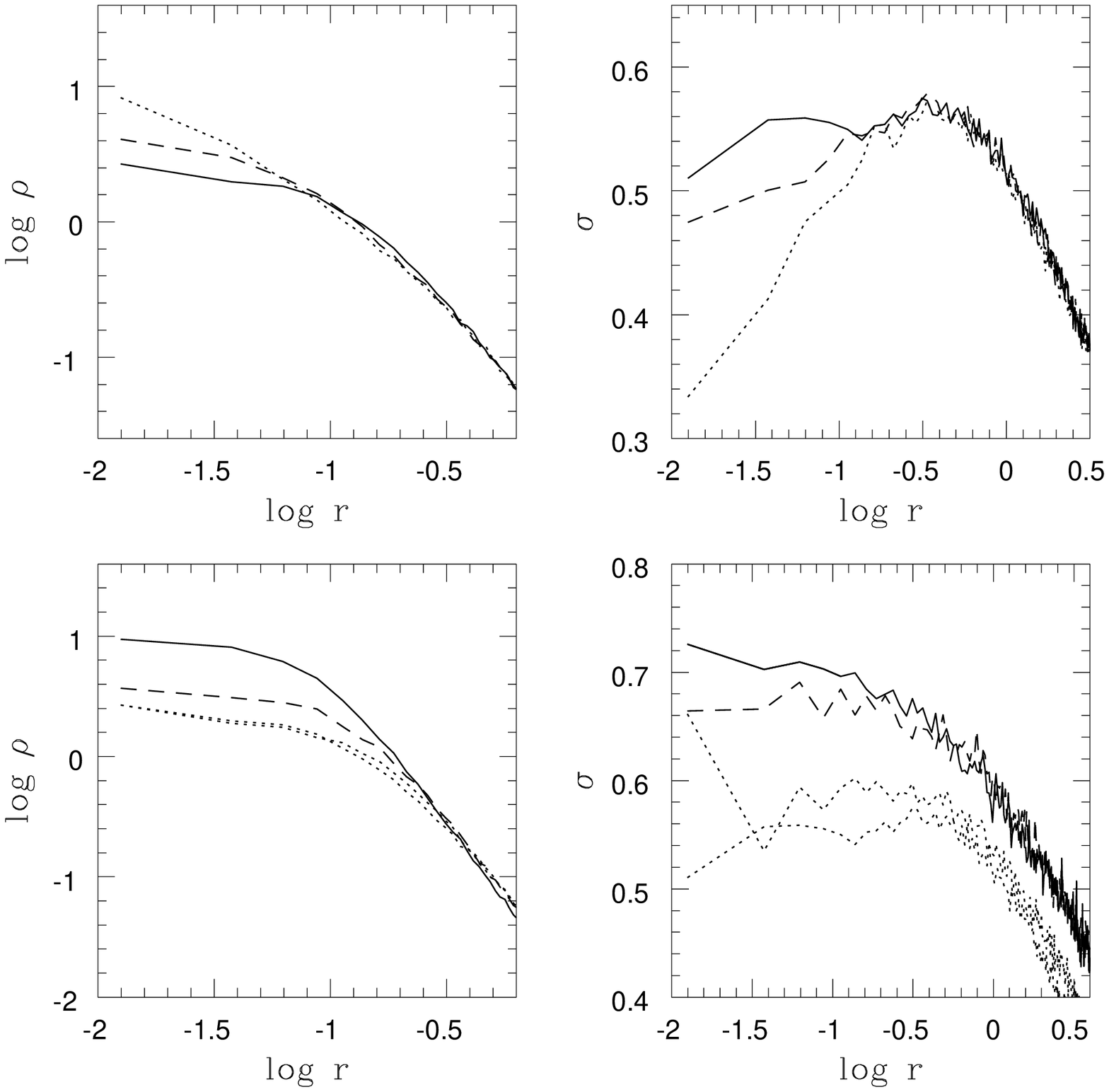,height=14cm}}
\caption{ Figure 1: The evolution of the dark matter density distribution $\rho$(r)
and of the 3-dimensional velocity dispersion profile $\sigma$(r)
in dimensionless units is shown,
adopting a weak interaction cross section of $\sigma^*$ = 10 r$_s^2$/M$_s$.
The upper panels show the phase of core expansion with the dotted curves
representing the initial conditions and the dashed and solid curves corresponding to
the evolutionary state of t=1 $\tau_{dyn}$  and t=2.5 $\tau_{dyn}$, respectively.
The lower panels show the epoch after core expansion (lower dotted lines: t=2.5
$\tau_{dyn}$, upper dotted lines: t=5 $\tau_{dyn}$) and the subsequent phase
of core collapse (dashed line: t=$10 \tau_{dyn}$, solid line: t=16 $\tau_{dyn}$).}
\end{figure}

\begin{figure}[p]
\centerline{\psfig{figure=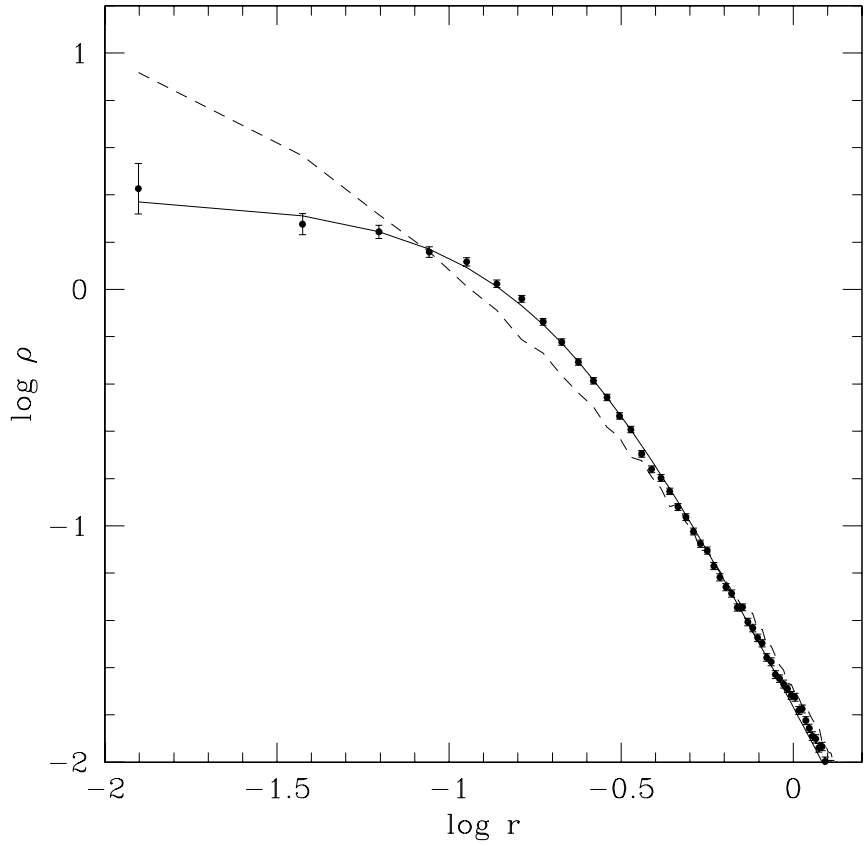,height=12cm}}
\caption{ Figure 2: The dark matter density distribution as inferred from 
rotation curves of dwarf galaxies (solid line) is compared with halo density
distribution at t=0 (dashed line ) and after core expansion at t = $\tau_{iso}$.}
\end{figure}
\end{document}